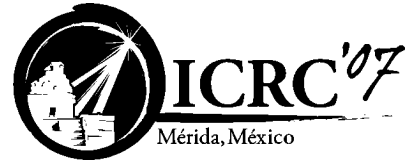

# Status report on project GRAND

J. POIRIER[1], C. D'ANDREA[1], E. FIDLER[1], J. GRESS[2], M. HERRERA[1], P. HEMPHILL[1], C. SWARTZENDRUBER[1]
[1]Physics Department, University of Notre Dame, Notre Dame, IN 46556 USA
[2]Productivity Management Inc., South Bend, IN 46628 USA
poirier@nd.edu

**Abstract:** GRAND is an array of position sensitive proportional wire chambers (PWCs) located at 86.2° W, 41.7° N at an elevation of 220 m adjacent to the campus of the University of Notre Dame with 82 square meters of total muon detector area. The geometry of the PWCs allows the angles of the charged secondary muon tracks to be measured to +/- 0.3 deg in each of two orthogonal planes. Muons are 99% differentiated from electrons by means of a 51 mm steel plate in each detector.

## The muon detector array: GRAND

Project GRAND is an air shower array of proportional wire chambers located at 41.7° N and 86.2° W at an elevation of 220 m above sea level and has been in operation with the construction of the complete 64 stations since 1995 (see Figure 1). Each station has an active area of 1.29 m² (for a total active muon detector area of 82 m²). Two individual triggers simultaneously take data for: 1) single muons (single tracks of identified muons) at ~2000 Hz and 2) extensive air showers (multiple stations hit in coincidence). The single track muon data are sensitive to primary energies >10 GeV with a median value of 56 GeV for vertical tracks. We concentrate in this paper on the power of GRAND as a muon detector. Additional details are in [3].

Project GRAND has demonstrated the ability to detect significant rises in ground level muon flux associated with solar flares [1,2] as well as the suppression in ground level muons associated with a Forbush decrease [3]. Project GRAND has also observed the daily variations in cosmic rays related to the interplanetary magnetic field (see Figure 2).

Since each plane has 80 signal wires, GRAND has the possibility of a large number of directional channels. These channels can be combined to achieve good statistics per channel as well as maintain a high angular resolution.

One such proposed method would combine the data into 15 x 15 = 225 directional channels each with a counting rate of approximately 30,000 muons / hour. This particular arrangement makes our data more easily compared to that of other muon detectors, but the fact that we start with angular data of 158 x 158 ~25,000 directional channels means that we can combine these narrow channels in various ways to match other muon detector's channel-widths thus making the data easier to compare as well as increasing the statistics in each of these wider angular channels. Data from Project GRAND can be automatically pressure corrected: $N=N_0\exp[1.2(P-P_0)/P_0]$ where N is the corrected rate, $N_0$ is the measured muon rate, P is the measured air pressure at that time, and $P_0$ is a reference pressure. These corrected muon rates can be made available in real time from pressure data which are locally measured and recorded.

Figure 2 shows these counting rate bins of one hour. It also shows a small daily variation which has been enhanced by showing only the top portion of the data. Even though the effect is small (deviations of about +/- 0.7%), the statistics of the muon counting rate makes these deviations significant. These deviations from a uniform muon secondary counting rate from cosmic ray primaries are presumably due to the quiet sun's Interplanetary Magnetic Field (IMF) on the galactic cosmic rays as observed by the secondary muons at ground level. Since these effects are distributed over a substantial portion of the IMF between the sun and the Earth, they have the potential of allowing us to study, on a daily basis, these effects of the solar wind and the IMF for a quiet sun in the absence of coronal mass ejections (CMEs). As we gain more information and understanding of these diurnal variations, they can be applied to forecasting the effects of CMEs arriving at the Earth from the sun. To the extent that the effect is due to the interplanetary magnetic field of the sun, a field which is not static and varies with time, a careful study of these variations with good statistics in the muon counting rate could give us a handle on these effects. It may be possible to get additional information by utilizing the angular data of the muons in addition to their absolute counting rate. Added details and publications for Project GRAND are available at: http://www.nd.edu/~grand.

## Forbush decrease of 29 October 2003

As an example of muon data on a Forbush decrease (FD), GRAND's muon data rate for October 29 and 30, 2003 is shown in the upper part of Figure 3. Additional data following the FD show that it takes about nine days to recover to its original counting rate. In order to ensure that the data reflect only physical variations, a cut was performed to select only the best stations for analysis for this event. Angular muon information was also obtained from October 29 to October 30. The average angle in the east-up direction and the north-up direction was calculated. The average angle in the north-up direction is shown in the lower portion of Figure 3. Since an FD is caused by the large coronal ejection which deflects incoming particles when it impacts the Earth, it should be expected that this would cause a deficiency in particles from a particular direction. In the data for the October 29, 2003 event, there is indeed a deficiency in southward originating particles occurring near the time of the sudden decrease.

## GLE of 20 January 2005

On 20 January 2005, NOAA reported an X7.1 X-ray solar flare beginning at 6:36 and peaking at 7:01 UT. LASCO reported an associated CME from the WNW-NW limb at 6:54 UT.

The 100 MeV protons at the GOES satellite peaked at 7:10. GRAND's muon counting rate was examined in three-minute bins from 5:00 to 10:00 UT. Twenty-eight stations which exhibited low rms deviations outside the signal region were selected for further analyses. The excess muon counting rate in the signal region from 6:51 to 6:57 UT was 7759 +/- 785 above a background of $3.72 \times 10^5$ muons. The angles of the muons in the signal region were examined, and their excess appeared to be limited to muon angles near vertical. Those muons whose angles lay from 13° S to 10° N and 10° W to 18° E were then selected for further analyses. In this rectangular box the possible excess muon rates above background are plotted in one-minute bins in Figure 4. In the time interval from 6:51 to 6:57 there is an excess of 5458 +/- 308 muons above a background of 74,400 muons. Since GRAND tags each muon with an absolute accuracy of two milliseconds, the time structure within this peak is planned to be studied in the data with more detail. (See e.g. [4].)

## Worldwide muon coverage

A single ground-based muon detector can only view a portion of the sky and, therefore, not always be in a good position on the Earth to collect data relevant to solar activity at a given time. A sufficiently dispersed worldwide network of muon detectors can monitor all regions of the sky at any given time and therefore be able to provide valuable information throughout the course of a solar storm. GRAND represents a muon station in North America with that capability and, together with stations in Japan, Brazil, Tasmania, Kuwait, and Germany, could make an important contribution to these muon detector stations distributed around the world.

## Education, archives, and outreach

Six Notre Dame undergraduate students have participated in research with GRAND and have written three theses. During the summers, undergraduate students from other universities are involved in an REU program, high school students in the REHS program, and high school teachers in the RET program. Their enthusiasm and quick assimilation to the basic analyses tools demonstrate the intuitive nature of the data from these PWC detectors and proves their capability to provide a good learning experience for budding scientists. The archiving of the data on accessible massive disk storage with the collaboration of Prof. D. Thain in the Engineering Department makes the data available to us as well as the high school students and teachers who have worked on the project to be able to do their own research on the data, which they helped collect.

The authors wish to thank: K. Neary for help with the manuscript; and the National Science Foundation for construction funds and their support of the REU program. Operating funds for GRAND are funded through the University of Notre Dame and private donations.

## References

[1] J. Poirier and C. D'Andrea. Ground level muons in coincidence with the solar flare of April 15, 2001. *Geophysical Research, Space Physics,* 107(A11):1376-1384, 2002.

[2] C. D'Andrea and J. Poirier. Ground level muons coincident with the 20 January 2005 Solar Flare. *Geophysical Research Letter*, 32:14102-14105, 2005.

[3] J. Poirier et al. A Study of the Forbush decrease of September 11, 2005 with GRAND. *30th International Cosmic Ray Conference,* paper #0985, 2007.

[4] H. Moraal, K.G. McCracken, P.H. Stoker. Analysis of the 20 January 2005 cosmic ray ground level enhancement. *30th International Cosmic Ray Conference,* paper #0862, 2007.

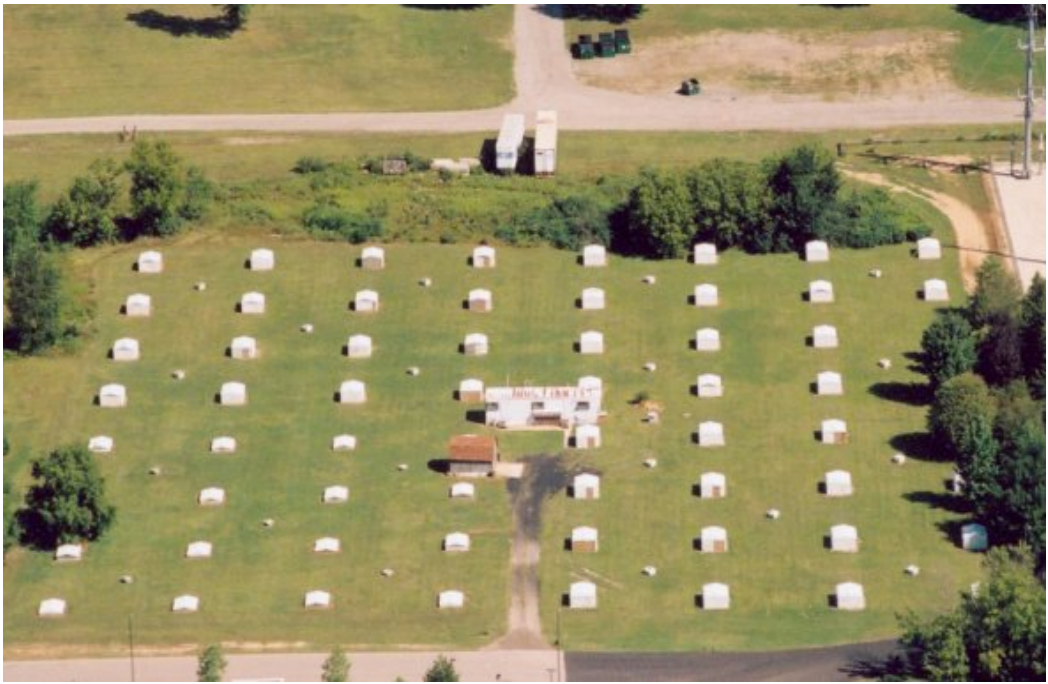

Figure 1: An aerial view of Project GRAND's muon detector stations.

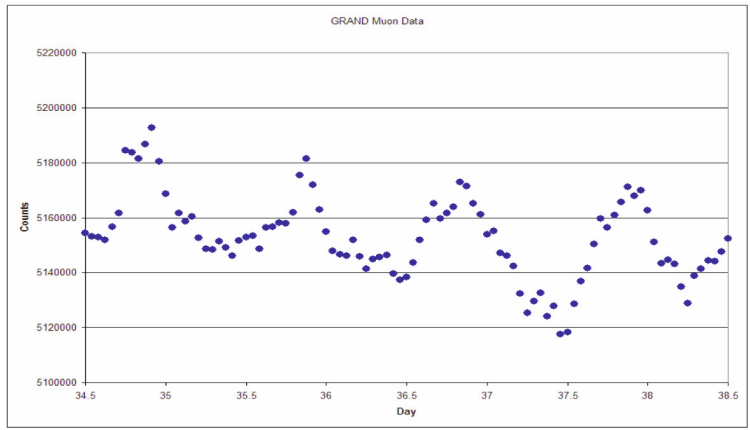

Figure 2: GRAND's pressure corrected counting rate versus time (EST). Notice the diurnal daily variations enhanced by the suppressed zero. These data will soon be available in real time.

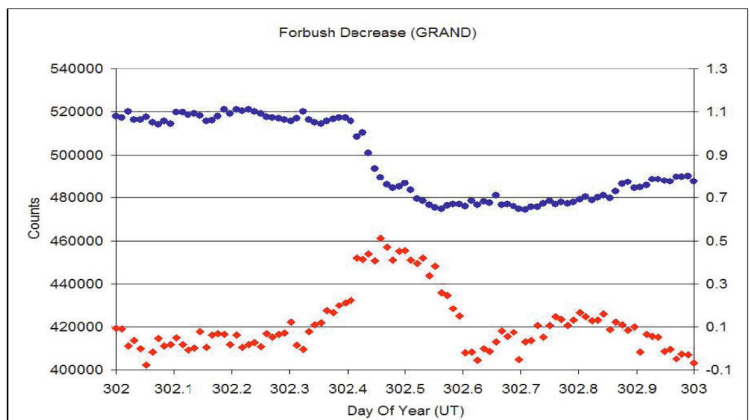

Figure 3: GRAND's muon counting rate (top points, left scale) and the mean angle of incoming muons in the north-up projected plane (bottom points, right scale). The angle changes during the FD onset.

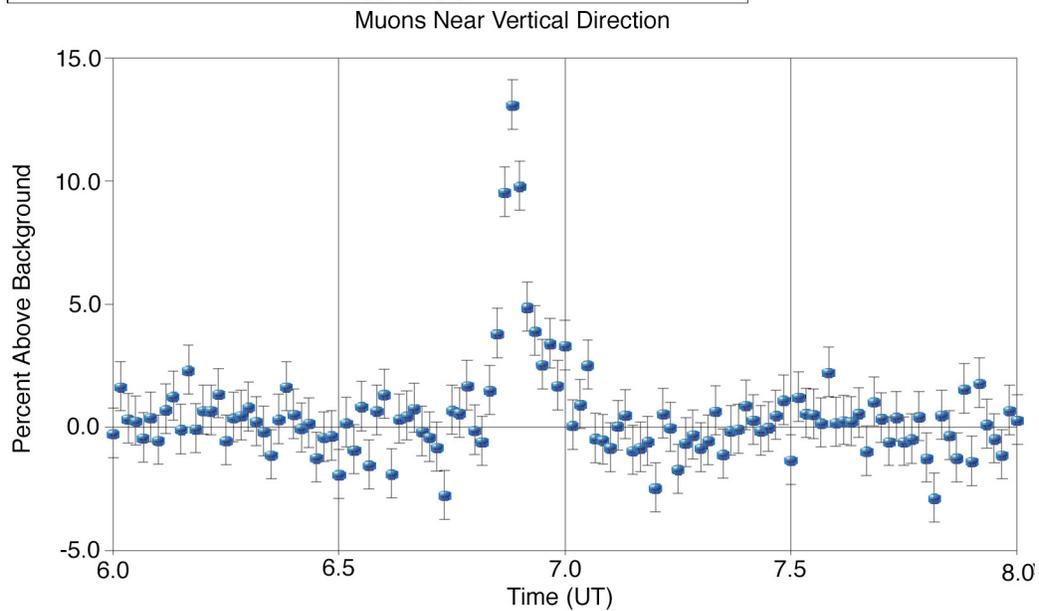

Figure 4. Muon rate in one-minute bins on 20 January 2005 for muons near vertical angle.